# Information transfer based on precision time synchronization via wireless interferometry

Daijiro Koyama[1], Yunzhuo Wang[1], Nobuyasu Shiga[2], Satoshi Yasuda[2], Nicolas Chauvet[1,3], and Makoto Naruse[1,3]

1. Department of Mathematical Engineering and Information Physics, Faculty of Engineering, The University of Tokyo, 7-3-1 Hongo, Bunkyo-ku, Tokyo 113-8656, Japan
2. Space-Time Standards Laboratory, Applied Electromagnetic Research Institute, National Institute of Information and Communication Technology, 4-2-1 Nukui-kitamachi, Koganei-shi, Tokyo 184-8795, Japan
3. Department of Information Physics and Computing, Graduate School of Information Science and Technology, The University of Tokyo, 7-3-1 Hongo, Bunkyo-ku, Tokyo 113-8656, Japan

**Abstract**

The growing demand of high-bandwidth and low-latency information transfer in information and communication technologies such as data centres and in-vehicle networks has increased the importance of optical communication networks in recent years. However, complicated arbitration schemes can impose significant overheads in data transfer, which may inhibit the full exploitation of the potential of optical interconnects. Herein, we propose an arbitration protocol based on precision time synchronization via wireless two-way interferometry (Wi-Wi), and numerically validate its efficiency including the ability to impose a strict upper bound on the latency of data transfer. Compared with the conventional carrier sense multiple access/collision detection (CSMA/CD)-based approach, a significant improvement in the data transfer was observed especially in the cases with high traffic flow rate. Furthermore, we conducted a proof-of-principle experiment for Wi-Wi-based data transfer between two electrically connected nodes and confirmed that the skew was less than 300 ns and remained stable over time. Conversely, non-WiWi-based data transfer exhibited huge and unstable skew. These results indicate that precision time synchronization is a promising resource to significantly reduce the communication overheads and ensure low latency for future networks and real-time applications.

## Introduction

With the growing importance of high-bandwidth and low-latency information transfer, optical communication technologies are expected to play a crucial role in short-distance communication networks such as data centres and in-vehicle networks in addition to conventional long-distance applications[1]. For example, self-driving cars need a variety of sensors such as light detection and ranging (LIDAR) sensor and stereo cameras as well as their controllers and associated computation modules, which generate a large amount of data. Hence, high-speed and low-latency communication is required in these cars[2]. In particular, the reduction of communication delay or latency is crucial for devices requiring strict real-time operation.

To realize the full potential of optical data transmission, the number of optical-to-electrical (OE) and electrical-to-optical (EO) conversions should be reduced as much as possible; otherwise, they can increase the delay and energy requirements. Further, arbitration is an important issue that requires careful attention, and our study focuses on this aspect of data transfer. To facilitate the removal of OE/EO conversions and to avoid unnecessary complexity, we have considered the case in which multiple nodes are connected in a unidirectional ring network. Since multiple nodes share access to a single bus, certain arbitration of communication is necessary to avoid collisions. Complex arbitration methods may impose significant overhead or delay, which can hinder the inherent advantages of optical signal transmission. Meanwhile, Shiga *et al.* developed a precision time synchronization technique called wireless two-way interferometry, also known as Wi-Wi[3-5] to address this issue. In this study, we have employed such a time synchronization technique to facilitate arbitration of communication in optical networks and even general information networks.

In particular, we have proposed a simple arbitration method for a ring network in which all the nodes are time synchronized via Wi-Wi, and demonstrate its efficiency including the ability to impose a strict upper bound on the latency of data transfer. Compared with the conventional carrier sense multiple access/collision detection (CSMA/CD) approach, a significant im- provement was observed in the proposed approach especially for the cases in which the traffic rate was high. Furthermore, we experimentally demonstrate the Wi-Wi-based data transfer between two electrically connected nodes and confirm the reduction in skew and its stability over time. We also show that the data transfer exhibits huge and unstable skew when Wi-Wi-based synchronization is turned off. These results pave the way for the realization of efficient and real-time communication by precise time synchronization for future applications.

## Principles

### Basic principle of time synchronization via Wi-Wi

Wi-Wi is a time synchronization technique and we employ modules that use 920-MHz wireless communication to achieve a time synchronization with a standard deviation of 20 ps[4]. Wi-Wi employs two-way time transfer (TWTT) technique[6], which is schematically shown in Fig. 1. Here, the time at sites A and B are denoted by $T_A$ and $T_B$, respectively, and the propagation delay between these sites is $\Delta$. To derive the time difference ($t_c = T_B - T_A$) and the propagation delay between the two sites, each site sends its time information via the same physical route. Then, the time difference at A when the information of $T_B$ is delivered is given by $\Delta_{atA} = T_A + \Delta - T_B = -t_c + \Delta$, while that at B is $\Delta_{atB} = t_c + \Delta$. Using these two differences, the time difference and propagation delay can be obtained as

$$t_c = (\Delta_{atB} - \Delta_{atA})/2, \tag{1}$$

$$\Delta = (\Delta_{atB} + \Delta_{atA})/2. \tag{2}$$

This allows us to synchronize the absolute time between the two sites.

TWTT was originally used for two-way satellite time and frequency transfer (TWSTFT), which is a time synchronization method based on geostationary satellites that is utilized in calibration of standard time[6,7]. TWSTFT achieves precise time synchronization with an accuracy of sub-picosecond by carrier-phase two-way satellite frequency transfer (TWCP)[7], which is a technique based on carrier phase to derive precise time difference. Wi-Wi is an extension of these techniques in which satellite communication is replaced with wireless communication to achieve precision time synchronization.

**Timing-based arbitration in ring network**

We consider a network in which multiple computing nodes are connected in a unidirectional ring structure and all the nodes are time synchronized via Wi-Wi, as shown in Fig. 2. It is assumed that add/drop of information between the nodes and the ring is technically achievable at each node. To describe the arbitration method based on timing, we introduce the notion of timing referred to as arbitration point (AP). AP is temporally transmitted in the ring regularly. Due to the time synchronization of all the nodes, each node can independently recognize AP. The duration of an AP is defined by "1 word".

Here we consider two states at AP: one is a 1-word-length non-zero-level signal defined by "1", while the other is a 1-word-length zero-level signal defined by "0". Each node can start sending the information content immediately after the time specified by AP. Here, we design the duration of the information content in such a way that the sum of total temporal length of AP and information content corresponds to the physical length of the ring. This implies that no more than two information contents can exist in the ring.

In addition, to ensure maximum latency among any given nodes, a node does not send information contents consecutively by utilizing two consecutive APs. This rule also ensures that all the nodes have an equal chance of data transmission. With these constraints, if the number of nodes in a ring is $n$ and the interval of AP is $T_{AP}$, the transmission queue length from a sender node to a receiver node can be strictly bounded by $nT_{AP}$. The flow of this arbitration is summarized in Algorithm 1.

---

**Algorithm 1:** Timing synchronization-based arbitration in a ring network

---

1. Detect the timing of AP at each node.
2. If a node wants to send information, i.e., the transmission queue contains certain message,
    1) Transmit "1" at AP.
    2) Simultaneously, if it receives "1" at AP from the ring, the node does not start transmitting information content to avoid a conflict. In addition, the node starts receiving the information content
       from the ring.
    3) When it detects "0" at AP from the ring, it starts transmitting the information content. Further, after finishing the transmission, "0" must be set in the next AP, i.e., the node cannot conduct consecutive data transmission.
3. If the node does not contain any information to be transmitted, i.e., the transmission queue is empty,
    1) Transmit "0" at AP.
    2) Simultaneously, if it receives "1" at AP from the ring, the node starts receiving the information content from the ring.
4. Return to step 1.

---

## Result and Discussion

### Performance analysis

We conducted simulations to compare the latency or delay obtained by conventional CSMA/CD-type approach and the proposed timing synchronization method. We evaluated two kinds of wait time. The first one is the difference between the time when a message enters the transmission queue and the time of its receipt by the receiver, which is called "Wait time 1". The second one is the difference between the arrival time of a message at the exit of the transmission queue and the time of its receipt by the receiver, which is called "Wait time 2".

Figures 3a and 3b show the Wait time 1 of CSMA/CD and the proposed method as a function of traffic rate, respectively. Each symbol in these figures represents the lower $x$ % ($x = 0, 10, 20,..., 100$) along with the minimum and maximum values of the wait time of all the transmitted messages with traffic rate ranging from 0.01 to 0.99. Figure 3a shows that when the traffic rate is more than 0.2 in CSMA/CD, the wait time increases drastically. When the traffic rate is less than 0.2, most of the messages are transmitted almost instantaneously, but some of messages wait for more than several tens of messages. On the other hand, as shown in Fig. 3b, in the proposed method, even if the traffic rate is nearly 0.8, nearly 90 % of the messages are transmitted within 10 messages, indicating that the proposed method maintains good delay tolerance even when the traffic rate is high.

Figures 3c and 3d summarize the distributions of Wait time 2 obtained by CSMA/CD and the proposed method, respectively, based on the analysis of 1,500 messages for each traffic rate. Occasionally, the Wait time 2 can be unexpectedly long in CSMA/CD, while it is strictly bounded by maximum 1050 words in the proposed method. Further, it is evident from Fig. 3d that almost all the messages are sent within 8 or 9 cycles, and very few messages wait for 8 or 9 cycles in the proposed method even when the traffic rate is 0.9 or higher.

Figure 4a shows the average of Wait time 2 of all the messages for each traffic rate. On average, the Wait time 2 of the proposed method is shorter than that of CSMA/CD regardless of the traffic rates. Meanwhile, even if the traffic rate is high, the average wait time in CSMA/CD is actually not very long. Therefore, it can be inferred that few messages which wait for extremely long time lead to a dramatic increase in wait time. This is clearly observed in Fig. 3a. According to Fig. 4b, the maximum wait time in CSMA/CD exponentially increases with the increase in the traffic rate, which severely impacts the scenarios in which real-time operation or short latency is crucial. Conversely, the proposed method provides a decrease of nearly $10^3$ in the maximum delay when the traffic rate is high.

**Experimental evaluation of skew using Wi-Wi**

The proposed method is based on precision time synchronization in data transmission. We experimentally evaluated the skew of information transmission using Wi-Wi to examine the validity of one of the most fundamental aspects of the proposed arbitration method. Figure 5a shows a schematic of the experimental setup, and Fig. 5b shows a snapshot of the experimental apparatus where two nodes, i.e., node A and B, are interconnected while their timings are synchronized via Wi-Wi. Each node sends messages to the other node and to itself via the same route. A node starts sending a message when it is triggered by the edge of the pulse generated by Wi-Wi with a rate of 1 pulse per second (pps). The 1-pps trigger signals on both the nodes are synchronized after Wi-Wi is switched on. The messages are transmitted at the rate of 20 Mbps, which corresponds to the base clock signals (10 MHz) supplied by Wi-Wi. This indicates that the time duration of a single bit corresponds to 50 ns. The details of the experimental setup are described in the Methods section.

We measured the skew between the first signals when Wi-Wi was activated and deactivated. In each case, we measured the skew nearly every 6 s to examine its time variation. The skew is defined as follows: arrival time of message from node B – arrival time of message from node A. As shown in Fig. 5c, by activating Wi-Wi, the skew between the first signals can be reduced to nearly 1/100 with respect to its value in the case where Wi-Wi is deactivated. On the contrary, even when the time synchronization is activated, a skew of at most 6 bits or 300 ns is observed. However, Fig. 6b shows that the skew can be maintained within ±1 bit of its initial value for about 10 min by activating Wi-Wi, but it increases by 300 bit when Wi-Wi is deactivated, as shown in Fig. 6a. This implies that the skew can be controlled within 1 bit by offset adjustment, which is necessary for the proposed arbitration protocol to recognize the timing of message transmission.

For comparison, we measured the skew between the nodes when their timing information were calibrated by connecting them to a network time protocol (NTP) server prior to the measurements. (The details of NTP is described in the Methods section.) Specifically, the connection to the NTP server was made before the measurement, and each node sent a message every second based on its own internal reference clock. Figures 6c and 6d show the time evolutions of skew in two independent measurements. Note that the unit of skew is ms, not bit. First, if we focus on the skew at the beginning of the measurement, it is clear that the skew is in the ms range in both Fig. 6c and 6d, even immediately after the connections to the NTP server. This skew is 5–6 orders of magnitude larger than that in the case based on Wi-Wi. In addition, the slope of the graph is equal in these two figures, and the skew is reduced by approximately 2.5 ms per 10 min. This indicates that the individual device-dependent variations, such as that due to reference clocks inside the two PCs and/or digital I/O devices, lead to timing variations. From this particular experiment, we observe that the reference clock on Node B is faster than that on Node A. These results indicate that not only the accuracy of the time synchronization by NTP is insufficient, but also that the time synchronization is easily lost due to the uneven performance of the reference clock between nodes.

**Future scope**

First, in the present study, we have assumed a ring-structure network due to its simplicity and high technological relevance to the optical ring network studied in the existing literature[8,9]. However, other network structures such as tree, star, and their combinations may be interesting for future investigations.

In the present study, the packet rate of information transmission in the experiment is defined through the trigger signal supplied by the Wi-Wi module, which is 1 pps. However, this rate can be much higher in practical applications. In future, we hope to address this issue using technological solutions including the latest field-programmable gate-array (FPGA) technology.

## Conclusion

We demonstrated the significance of reducing arbitration overhead and ensuring information-transfer latency to realize the full potential of high-bandwidth optical communications and to facilitate data transfer applications in which real-time operation or short latency is critical. We proposed an arbitration protocol in a ring-structure network based on precision time synchronization via Wi-Wi, which ensured that the latency of data transfer was strictly bounded by an upper limit, while conventional CSMA/CD resulted in a very high latency especially in the cases with high traffic rate. Furthermore, we conducted a proof-of-principle experiment for data transfer between two electrically connected nodes that were time synchronized via Wi-Wi. We confirmed that the skew was not more than 300 ns and remained stable over time. Conversely, huge and unstable skew was observed in the absence of time synchronization. These results indicate that precise time synchronization is a vital resource to significantly reduce the communication overheads and to ensure that the latency is smaller than the pre-defined value for future networks and real-time applications.

## Methods

**Simulation settings in delay analysis**

To simulate short-distance, small-scale communication networks such as data centres and in-vehicle networks, the following settings were considered:

- The bus length is equivalent to $L = 100$ words.

- The message size is $M = 105$ words (equivalent to the interval between consecutive APs for the time synchronization method. For CSMA/CD, it is equivalent to 106 words because the last word of transmission is included.)
- The number of nodes is $N = 10$ and they are randomly placed.
- $R$ is the traffic rate ranging from 0 to 1, and a message is generated at each node with the same probability of $R/(NM)$ per word and is added to the transmission queue.
- The simulation time is equivalent to $10^7$ words.

The details of the CSMA/CD algorithm are as follows:

1. When a node receives the header of a message addressed to itself
    (a) Receive data for the message size.
    (b) When the last word of transmission is received after the message size, the received message is treated as a valid message.

2. When there is a message in the transmission queue, and the node is not waiting.
    (a) Start to send the message.
    (b) Receive the downstream data while transmitting, and check whether it is same as the transmitted message.
    (c) If the transmitted message is different from the received downstream data, a jam signal is sent to notify the other nodes that a collision has occurred.
    (d) Assuming that the number of times this message has collided is $c$, wait for the size of $[0, 2^{\min(10,c)})$ messages.

**Experimental scheme for skew evaluation**

Wi-Wi modules use 920-MHz wireless signal for time synchronization. The digital I/O devices used in the experiment were PCIe-6537B manufactured by National Instruments, and its main specifications are summarized in Table 1. The digital I/O devices were connected to interface board CB-2162 via the cable 195949A-01_C68-C68-D4. The digital I/O devices were controlled by LabVIEW 2019. The length of the wires used to connect the interface boards was 20 cm, and the length of the wires between Wi-Wi and the interface boards was approximately 53 cm. A message included 37 digits of 0 and 1 data. The specifications of PC are summarized in Table 2. The NTP server (ntp.nict.jp) was provided by the National Institute of Information and Communications Technology. This server is located in Nukui-Kitamachi, Koganei, Tokyo. The connection with the NTP server was achieved using

Windows built-in time setting.

**Data availability**

The datasets generated during the current study are available from the corresponding author on reasonable request.

**Acknowledgements**

This work was supported in part by the CREST project (JPMJCR17N2) funded by the Japan Science and Technology Agency and Grants-in-Aid for Scientific Research (JP17H01277, JP20H00233) funded by the Japan Society for the Promotion of Science.

**Author contributions**

M.N. directed the project. Y.W., D.K., N.S., S.Y., N.C., and M.N. conceptualized the research. N.S. and S.Y. developed the Wi-Wi technology. D.K., N.C., and M.N. designed and implemented the experiments. D.K. conducted the measurements. D.K., N.S., S.Y., N.C., and M.N. analyzed the data. D.K., N.C. and M.N wrote the paper. All the authors provided critical feedback and approved the final manuscript.

**Competing interests**

The authors declare no competing interests.

**Additional information**

Correspondence and requests for materials should be addressed to D.K. and M.N.

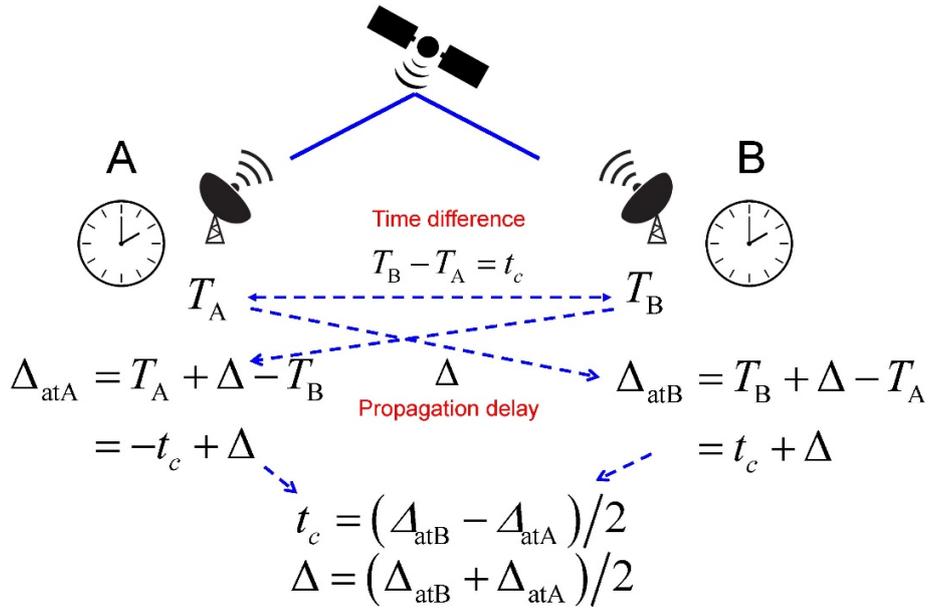

**Figure 1.** Schematic of two-way time transfer technique.

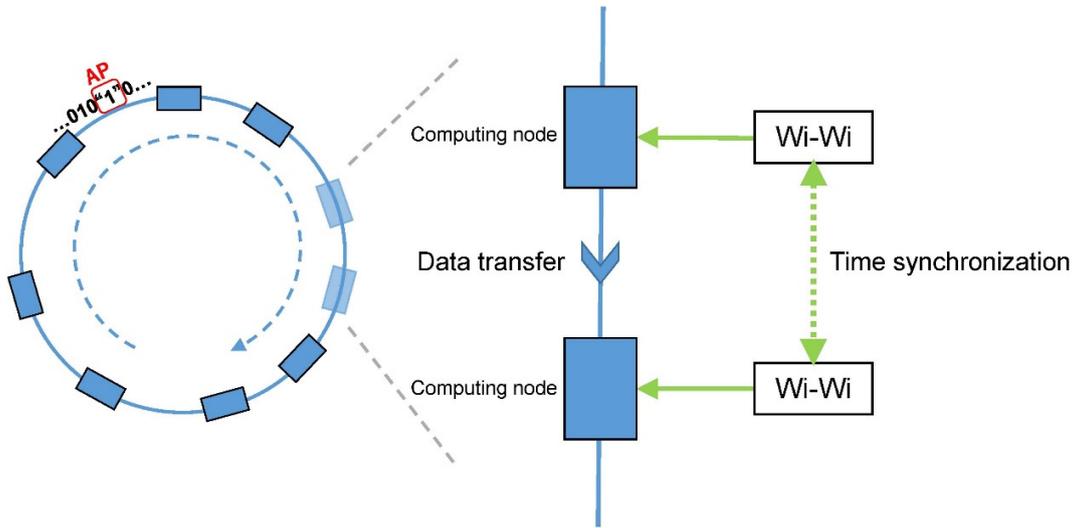

**Figure 2.** Multiple computing nodes are connected in an unidirectional ring network where all nodes are synchronized via Wi-Wi.

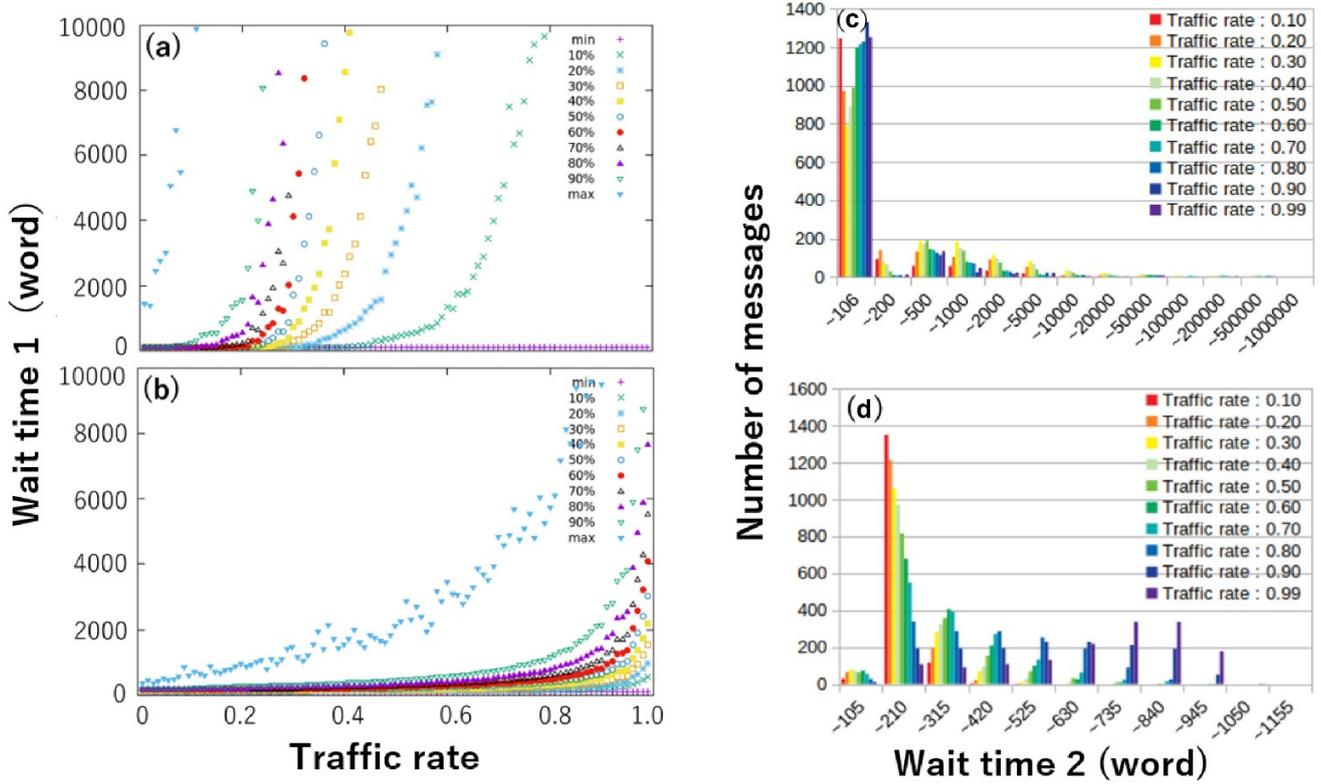

**Figure 3.** Latency analysis by the proposed method. **(a, b)** Wait time 1 obtained by CSMA/CD (a) and proposed method (b). **(c, d)** Wait time 2 obtained by CSMA/CD (c) and proposed method (d).

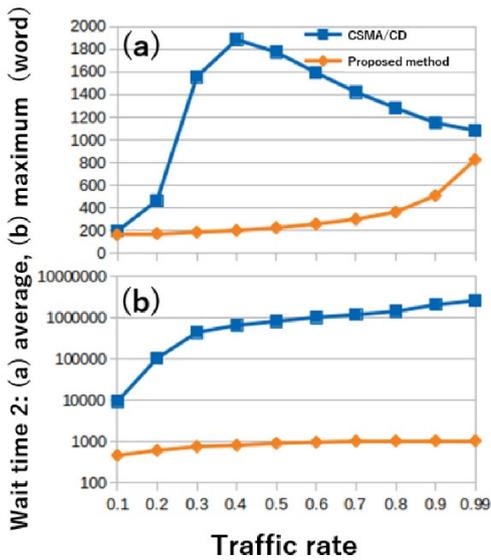

**Figure 4.** Comparison of **(a)** the average and **(b)** the maximum value of Wait time 2 obtained by CSMA/CD and the proposed method.

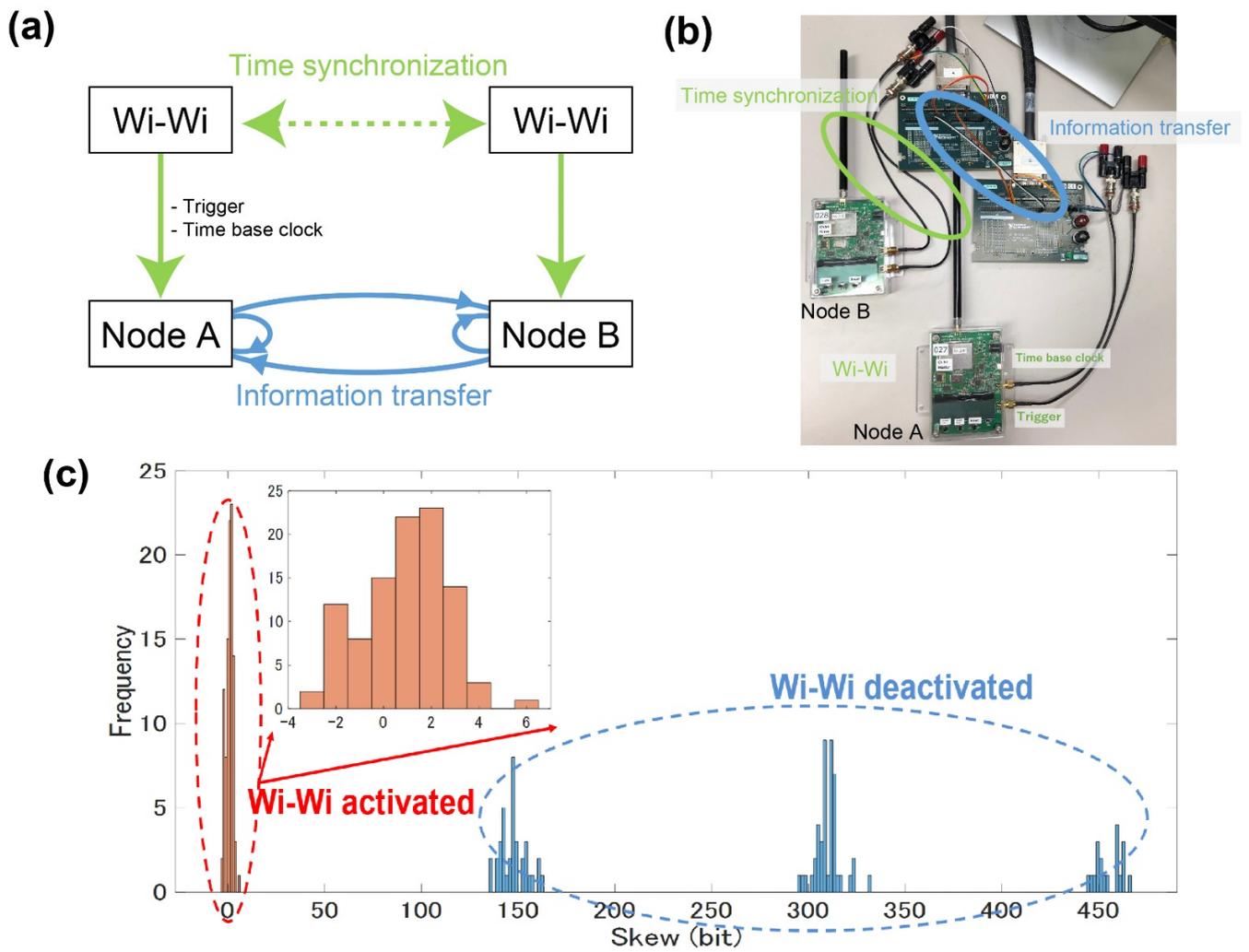

**Figure 5.** Experimental demonstrations. **(a)** Schematic of the experimental setup. **(b)** Overview of the experimental setup. Each interface board is connected to the input/output (I/O) devices in the corresponding PC via the upper black wire. **(c)** Skew between the first signals when Wi-Wi is activated and deactivated. We conducted 100 measurements for each case.

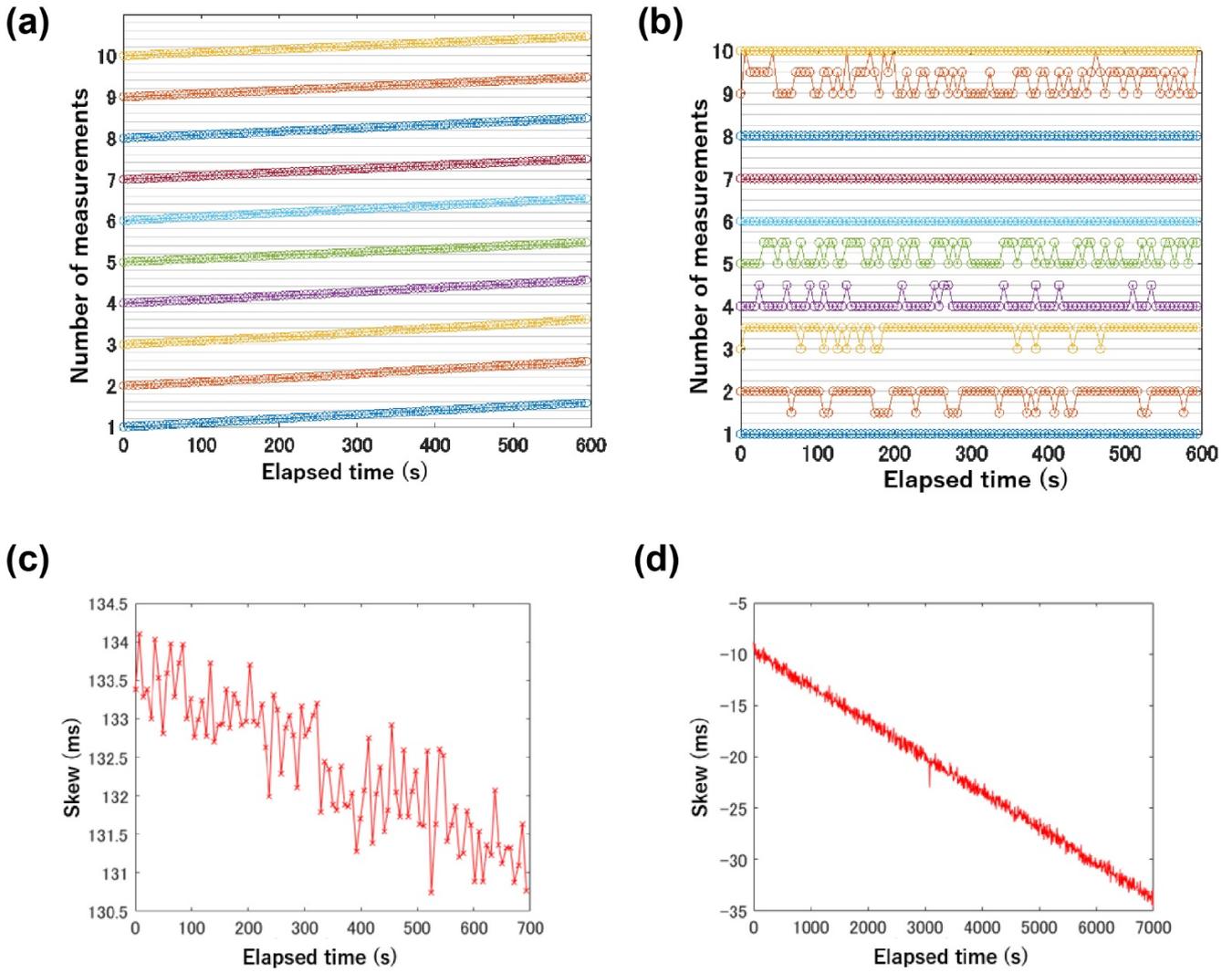

**Figure 6.** Stability and instability of skew with and without Wi-Wi. **(a)** Ten measurements for the time variation of skew when Wi-Wi is deactivated. The interval of grid lines is equivalent to 100 bits. **(b)** Ten measurements for the time variation of skew when Wi-Wi is activated. Two intervals of grid lines are equivalent to 1 bit. **(c,d)** Time variation of skew when the nodes are connected to the NTP server before measurements. We performed (c) 100 and (d) 1000 measurements every 7 s. In **(d)**, 13 points are omitted due to data loss.

**Table 1.** Representative specifications of the interface used in the experiment (PCIe-6537B)

| Specification | Value |
|---|---|
| Sample clock timebase | 200 MHz internal oscillator |
| Sample clock | onboard clock (sample clock timebase with divider), imported sample clock |
| Onboard clock frequency range | 200 MHz / $N$ ($4 \leq N \leq 4,194,307$) |
| Imported sample clock frequency range | 0 Hz – 50 MHz |
| Onboard memory | 2,048 sample |
| Delay from trigger to digital data output | 65 ns to 200 ns (1 sample clock cycle + 150 ns) |

**Table 2.** Specifications of PCs used in the experiment

| Specification | PC for node A | PC for node B |
|---|---|---|
| Model number | Dell Precision 5820 Tower | Dell Precision Tower 3620 |
| OS | Windows 10 Pro for workstations (64 bit) | Windows 10 Pro (64 bit) |
| OS version | 1903 | 1903 |
| CPU | Intel® Xeon® W-2102, 2.90 GHz | Intel® Core™i5-6500, 3.20 GHz |
| RAM | 16.0 GB | 8.00 GB |